\documentclass[superscriptaddress, reprint, amsmath, amssymb, aps, pra, floatfix]{revtex4-2}

\usepackage{fullpage} %
\usepackage{parskip} 
\usepackage{tikz} 

\usepackage{mathtools}
\usepackage{bm}
\usepackage{physics}
\usepackage{amsfonts}
\usepackage{fancyhdr}
\usepackage{times}
\usepackage{changepage}
\usepackage{amssymb}
\usepackage{appendix}
\usepackage{amsthm}
\usepackage[english]{babel}
\usepackage{graphicx}
\usepackage{xcolor,colortbl}
\usepackage{amsmath}
\usepackage{lipsum}
\usepackage{wrapfig}
\usepackage{float}
\usepackage{color} 
\usepackage{xcolor}
\usepackage{bbold}
\usepackage{hyperref}
\hypersetup{colorlinks = true}

\begin{document}

\title{Quantum tracking control of the orientation of symmetric top molecules}

\author{Alicia B. Magann}
\email{abmagan@sandia.gov}
\affiliation{Department of Chemical \& Biological Engineering, Princeton University, Princeton, New Jersey 08544, USA}
\affiliation{Center for Computing Research, Sandia National Laboratories, Albuquerque, New Mexico 87185, USA}

 \author{Tak-San Ho}
 
 \affiliation{Department of Chemistry, Princeton University, Princeton, New Jersey 08544, USA}
 
 \author{Christian Arenz} 
  \affiliation{Department of Chemistry, Princeton University, Princeton, New Jersey 08544, USA}
    \affiliation{School of Electrical, Computer and Energy Engineering, Arizona State University, Tempe, Arizona 85281, USA}

 \author{Herschel A. Rabitz}
 \email{hrabitz@princeton.edu}
  \affiliation{Department of Chemistry, Princeton University, Princeton, New Jersey 08544, USA}

\date{\today}

\begin{abstract}

The goal of quantum tracking control is to identify shaped fields to steer observable expectation values along designated time-dependent tracks. The fields are determined via an iteration-free procedure, which is based on inverting the underlying dynamical equations governing the controlled observables. In this article, we generalize the ideas in {Phys. Rev. A 98, 043429 (2018)} to the task of orienting symmetric top molecules in 3D. To this end, we derive equations for the control fields capable of directly tracking the expected value of the 3D dipole orientation vector along a desired path in time. We show this framework can be utilized for tracking the orientation of linear molecules as well, and present numerical illustrations of these principles for symmetric top tracking control problems. 

\end{abstract}

\maketitle

\section{Introduction}

The desire to selectively manipulate molecular dynamics using external fields is a decades-old dream that has motivated a broad range of research pursuits \cite{Brif2011,Glaser2015,0953-8984-28-21-213001}, including the development of quantum optimal control (QOC) theory \cite{Peirce1988}. The goal of QOC is to identify fields to control the dynamics of a quantum system, such that the system achieves a desired control objective at a designated target time $t=T$. The task of identifying an optimal field is typically accomplished by iterative optimization methods \cite{PhysRevA.84.022326,doi:10.1063/1.1564043,doi:10.1063/1.476575}. Although these methods can be computationally demanding, QOC has nonetheless found broad applications, ranging from quantum computing \cite{Dolde2014,Waldherr2014} to chemical reactions \cite{Assion919, Levis709, 2002EPJD...20...71D, PhysRevLett.94.068305}.

In this article, we focus on another formulation, quantum tracking control (QTC) \cite{Gross1993,Chen1995,Chen1997}, for designing control fields to accurately track the temporal path of an observable of interest. The origins of QTC are in engineering control theory, which has explored tracking control in a range of settings including linear \cite{BROCKETT1965548}, nonlinear \cite{Hirschorn1979}, and bilinear \cite{Ong1984} systems. For quantum-mechanical applications, tracking control principles have been applied towards the numerical study of systems including a qubit \cite{Lidar2004}, a single atom \cite{PhysRevLett.118.083201}, and various molecular \cite{Gross1993,Chen1995,Chen1997,PhysRevA.98.043429,magann2020scalable} and solid-state systems \cite{PhysRevA.101.053408,PhysRevLett.124.183201,mccaul2021optical}. 

\begin{figure}[t] 
\centering
\includegraphics[width=0.75\columnwidth]{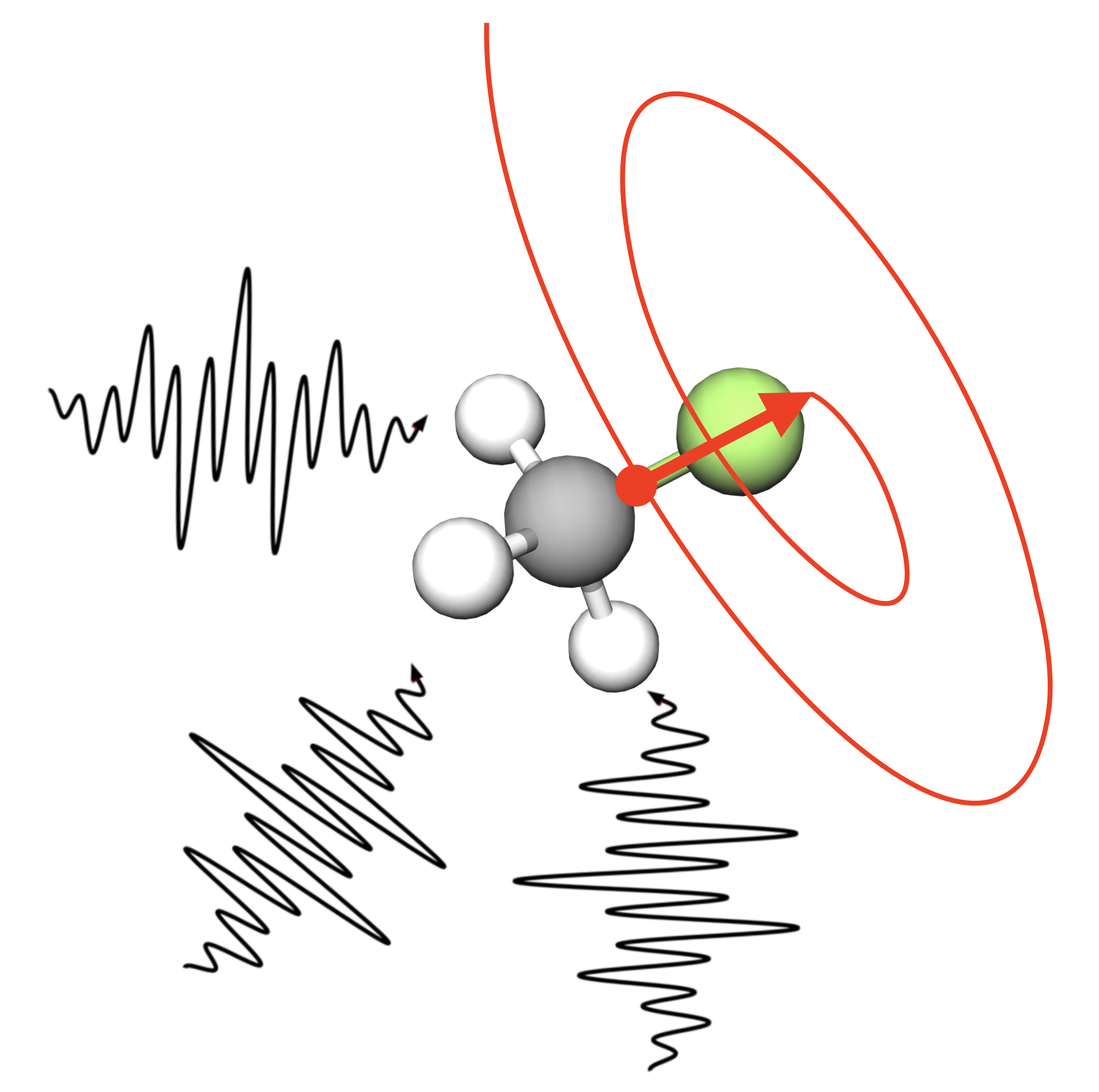}
\caption{In this article, we formulate QTC for controlling the 3D orientation of symmetric top molecules, such as fluoromethane, shown here. The control procedure involves designing three orthogonal fields (black) in order to drive the molecule's 3D dipole vector along a desired time-dependent track (red). The three fields can be determined by solving an inverse equation, without the need for optimization.}
\label{Fig:Intro}
\end{figure}

The aim of QTC is to find tracking control field(s) $\bm{\varepsilon}(t)$ that drive one or multiple observable expectation values $\langle \bm{O}\rangle(t)\equiv \langle\psi(t)|\bm{O}|\psi(t)\rangle$ along desired time-dependent ``tracks'' $\langle \bm{O}\rangle_d(t)$ for a chosen time interval $t\in[0,T]$. This is carried out by directly inverting the underlying dynamical equation governing $\langle \bm{O}\rangle(t)$ in order to solve for $\bm{\varepsilon}(t)$ \cite{Gross1993,Chen1995,Chen1997}. Because it does not require any iterative optimization, QTC can be computationally advantageous compared with usual QOC schemes. 

A challenge facing QTC is the potential presence of singularities in the corresponding direct inversion procedure \cite{Zhu1905}. That is, attempts to exactly track arbitrary time-dependent observable paths can produce unphysical, discontinuous control fields \cite{Hirschornf1987} and deviations from the desired tracks. However, if singularities can be avoided, QTC offers an appealing, iteration-free approach for designing fields to control quantum systems.

Here, we consider applications of QTC to orienting symmetric top molecules. The control of molecular orientation has applications spanning high harmonic generation \cite{PhysRevLett.109.233903} and chemical reaction enhancement \cite{Brooks11,Zare1875,Rakitzis1852}, and has been the subject of numerous experimental \cite{exp3,exp2,exp1} and theoretical \cite{hoki2001,salomon2005,Turinici2010,yoshida2015,Yu2017, szidarovszky2018limao,ma2020optimal} studies. In particular, QTC of molecular rotor orientation in 2D has been explored \cite{PhysRevA.98.043429}. In this work, we extend this prior work to linear and symmetric top molecules in 3D. We note that although the controllability of linear and symmetric top molecules has been the subject of other studies \cite{boscain2014multi,2019arXiv191001924B}, to the best of our knowledge, the suitability of symmetric tops for QTC has thus far not been explored. 

The remainder of the paper is organized as follows. We begin by outlining the symmetric top rotor model and derive QTC equations for the control fields to track its orientation. We go on to describe computational methods for solving the QTC equations by expanding the wave function in terms of angular momentum eigenfunctions of the symmetric top and address the QTC singularity issue. We then show how the formulation of QTC for symmetric top molecules can be reduced to the case of a linear rotor. We conclude with numerical illustrations and an outlook.

\section{Symmetric top molecules in 3D}\label{Sec:3qtcfields}

We consider a symmetric top molecule with dynamics governed by the time-dependent Schr\"odinger equation,
\begin{equation}
    i\frac{\partial}{\partial t}|\psi(t)\rangle = H(t)|\psi(t)\rangle\, ,
    \label{eq:3d1}
\end{equation}
where $\hbar=1$ and the time-dependent Hamiltonian is
\begin{equation}
    H(t) = H_0-\bm\mu\cdot\bm\varepsilon(t) 
    \label{Eq:TheHamiltonian}
\end{equation}
in terms of (1) the field-free Hamiltonian $H_0$, (2) three orthogonal control fields $\varepsilon_X(t)$, $\varepsilon_Y(t)$, and $\varepsilon_Z(t)$, i.e., where $\bm{\varepsilon}(t)=\hat X\varepsilon_X(t)+\hat Y\varepsilon_Y(t)+\hat Z\varepsilon_Z(t)$, and (3) the components of the dipole moment $\bm{\mu} = \hat X\mu_X+\hat Y\mu_Y+\hat Z\mu_Z$, where $\hat X$, $\hat Y$, and $\hat Z$ denote the three Cartesian unit vectors in the laboratory, space-fixed frame of reference. 

\begin{figure}[b!] 
\centering
\includegraphics[width=0.6\columnwidth]{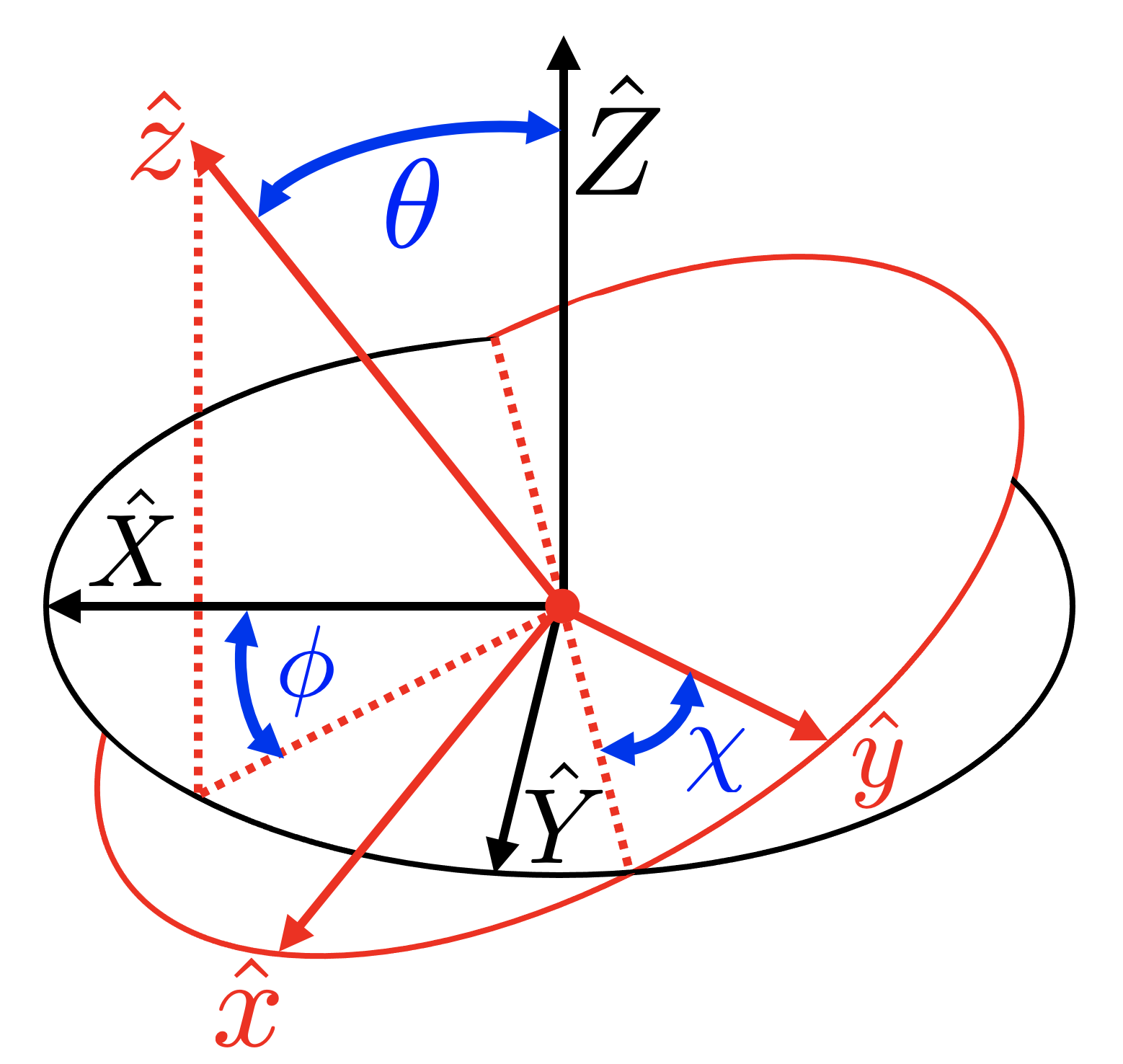}
\caption{Diagram showing $(\theta,\phi,\chi)$ Euler angle relations between laboratory space-fixed $\hat X$, $\hat Y$, and $\hat Z$ coordinates (black) and molecular body-fixed $\hat x$, $\hat y$, and $\hat z$ coordinates (red).}
\label{Fig:EulerAngles}
\end{figure}

Given the symmetry of the molecule, the dipole moment is along the principal, molecular, body-fixed $\hat{z}$-axis, such that $\bm \mu = \hat z \mu_z$, where $\mu_z = \mu z$, $\mu$ is the magnitude of the dipole moment, and $z$ is the body-fixed position operator. Noting that vectors represented in body-fixed coordinates $\hat x$, $\hat y$, and $\hat z$ and space-fixed coordinates $\hat X$, $\hat Y$, and $\hat Z$ can be related via Euler angles $\theta\in[0,\pi]$, $\phi\in[0,2\pi]$, and $\chi\in [0,2\pi]$, as per Fig. \ref{Fig:EulerAngles}, the components of the dipole moment in the space-fixed frame are then given by
\begin{equation}
\begin{aligned}
    \mu_X &= \mu  X = \mu \sin\theta\cos\phi,\\
    \mu_Y &= \mu Y  = \mu\sin\theta\sin\phi,\\
    \mu_Z &= \mu Z  = \mu\cos\theta,
    \label{Eq:dipoles}
    \end{aligned}
\end{equation}
where $X,Y,Z$ denote the space-fixed position operators, expressed using Euler angles $\theta,\phi$.

The molecule is assumed to be a rigid rotor, and the field-free symmetric top Hamiltonian is given by \cite{Zare}
\begin{equation}
    H_0=B(J_x^2+J_y^2)+CJ_z^2,
    \label{Eq:HintermsofJs}
\end{equation}
where $B$ and $C$ are rotational constants and $J_x$, $J_y$, and $J_z$, respectively, denote angular momentum projection operators in the molecular frame, given by the relations
\begin{equation}
    \begin{aligned}
    J_x&=-i\cos\chi\bigg(\cot\theta\frac{\partial}{\partial\chi}-\frac{1}{\sin\theta}\frac{\partial}{\partial\phi}\bigg)-i\sin\chi\frac{\partial}{\partial\theta},\\
    J_y&=i\sin\chi\bigg(\cot\theta\frac{\partial}{\partial\chi}-\frac{1}{\sin\theta}\frac{\partial}{\partial\phi}\bigg)-i\cos\chi\frac{\partial}{\partial\theta},
    \end{aligned}
    \label{Eq:Jxy}
\end{equation}
and
\begin{equation}
    J_z=-i\frac{\partial}{\partial\chi}.
    \label{Eq:Jz}
\end{equation}
As a result, the total angular momentum can be written as
\begin{equation}
    \begin{aligned}
    \mathbf{J}^2 &= -\Bigg(\frac{\partial^2}{\partial\theta^2}+\cot\theta\frac{\partial}{\partial\theta}+\frac{1}{\sin^2\theta}\bigg(\frac{\partial^2}{\partial\phi^2}+\frac{\partial^2}{\partial\chi^2}\\
    &\quad -2\cos\theta\frac{\partial^2}{\partial\phi\partial\chi}\bigg)\Bigg)
    \end{aligned}
\end{equation}
and the field-free Hamiltonian becomes
\begin{equation}
\begin{aligned}
    H_0 &= -B\bigg(\frac{\partial^2}{\partial\theta^2}+\cot\theta\frac{\partial}{\partial\theta}+\cot^2\theta\frac{\partial^2}{\partial\chi^2}\\
    &\quad-2\frac{\cot\theta}{\sin\theta}\frac{\partial^2}{\partial\chi\partial\phi}+\frac{1}{\sin^2\theta}\frac{\partial^2}{\partial\phi^2}\bigg)-C\frac{\partial^2}{\partial\chi^2}.
    \end{aligned}
    \label{eq:symtoph0}
\end{equation}

\section{Quantum tracking control equations for symmetric top orientation}\label{sec3}

Here, we apply the QTC framework \cite{Gross1993,Chen1995,Chen1997,PhysRevA.98.043429} to tracking a symmetric top molecule's 3D orientation using three orthogonal QTC fields. The time-dependent symmetric top orientation is defined as 
\begin{equation}
    \langle\mathbf R\rangle(t)=\hat X\langle X\rangle(t)+\hat Y \langle Y\rangle(t)+ \hat Z \langle Z\rangle(t),
\end{equation}
which is the instantaneous expectation value, at time $t$, of the position vector operator $\mathbf R\equiv \hat X X +\hat Y Y +\hat Z Z$. By differentiating $\langle \mathbf R \rangle(t)$ with respect to $t$ once we obtain
\begin{equation}
    \frac{d\langle \mathbf{R}\rangle(t)}{dt} = i\langle[H_{0},\mathbf{R}]\rangle(t),
    \label{Eq:firstoder}
\end{equation}
which has no explicit dependence on $\bm{\varepsilon}(t)$. By further differentiating Eq. (\ref{Eq:firstoder}) with respect to $t$ we obtain 
\begin{align}
 \label{dynamicssecondorder}
\frac{d^{2}\langle \mathbf{R}\rangle(t)}{dt^{2}}=\langle[\bm\mu\cdot\bm\varepsilon(t) ,[H_{0},\mathbf{R}]] \rangle(t) -\langle[H_{0},[H_{0},\mathbf{R}]]\rangle(t).
\end{align}
Eq. \eqref{dynamicssecondorder} can be expressed as a single matrix equation $\mathbf{b}(t)=\mathbf{A}(t)\bm{\varepsilon}(t)$, where $\bm{\varepsilon}(t)=(\varepsilon_{X}(t),\varepsilon_{Y}(t),\varepsilon_{Z}(t))^{T}$, the components of the matrix $\mathbf{A}(t)$ are given by
\begin{equation}
    \begin{aligned}
    \mathbf{A}_{X,X}(t) &= \langle[\mu_X,[H_0,X]]\rangle (t)= 2\mu B\langle Y^2+Z^2\rangle(t)\\ 
    \mathbf{A}_{Y,Y}(t)&=\langle[\mu_Y,[H_0,Y]]\rangle(t)=2\mu B\langle Z^2+X^2\rangle(t)\\
    \mathbf{A}_{Z,Z}(t)&=\langle[\mu_Z,[H_0,Z]]\rangle(t)=2\mu B\langle X^2+Y^2\rangle(t)\\
    \mathbf{A}_{X,Y}(t)  &= \mathbf{A}_{Y,X}(t) = \langle[\mu_Y,[H_0,X]]\rangle(t)\\
    &=-2\mu B\langle XY\rangle(t)\\
    \mathbf{A}_{Y,Z}(t)  &=\mathbf{A}_{Z,Y}(t) = \langle[\mu_Z,[H_0,Y]]\rangle(t)\\
    &=-2\mu B\langle YZ\rangle(t)\\
    \mathbf{A}_{Z,X}(t) & =\mathbf{A}_{X,Z}(t) = \langle [\mu_X,[H_0,Z]]\rangle(t)\\
    &= -2\mu B\langle ZX\rangle(t),\\
    \end{aligned}
    \label{ElementsofA}
\end{equation}
and the components of the vector $\mathbf{b}(t)$ read
\begin{equation}
    \mathbf{b}(t)=\frac{d^{2}\langle \mathbf{R}\rangle_{d}(t)}{dt^{2}}\langle +\langle[H_{0},[H_{0},\mathbf{R}]]\rangle(t).
    \label{bequation}
\end{equation}
Here the subscript “$d$” denotes the predefined or ``designated'' path in time to be tracked, $\langle \mathbf{R}\rangle_{d}(t)$.

The QTC fields can be found by inverting $\mathbf{A}(t)$, i.e., assuming the inverse of $\mathbf{A}(t)$ exists at all times $t$, 
and solving the resultant QTC equations, 
\begin{align}
\bm{\varepsilon}(t)=\mathbf{A}^{-1}(t)\mathbf{b}(t),
\label{QTCEquations}
\end{align}
as follows. First, the initial field values $\bm{\varepsilon}(0)$ are computed at time $t=0$, by evaluating \eqref{QTCEquations} for an initial state $|\psi(0)\rangle$. The next step is to evolve the system forward in time by integrating the Schr\"odinger equation \eqref{eq:3d1} over a small time step $\Delta t$, where this evolution depends on $\bm{\varepsilon}(0)$. Then, the state that results from this forward propagation, $|\psi(\Delta t)\rangle$, can be substituted into Eq. \eqref{QTCEquations} to compute  $\bm{\varepsilon}(\Delta t)$ associated with time $t=\Delta t$. This procedure is then repeated for all remaining time steps, where each forward step $k-1\rightarrow k$ involves the following two computational steps \emph{(i)} and \emph{(ii)}: 
\begin{itemize}
    \item[\emph{(i)}] $|\psi(k\Delta t)\rangle = e^{-iH\big(\bm{\varepsilon}\big((k-1)\Delta t\big)\big)\Delta t}|\psi\big((k-1)\Delta t\big) \rangle$
    \item[\emph{(ii)}] $\bm{\varepsilon}(k\Delta t) = \mathbf{A}^{-1}\big(|\psi(k\Delta t)\rangle\big)\mathbf{b}\big(|\psi(k\Delta t)\rangle\big)$.
\end{itemize}
The computational details associated with steps \emph{(i)} and \emph{(ii)} are given in Sec. \ref{seciv}. As mentioned above, this procedure requires that $\mathbf{A}(t)$ is invertible at all times. A singularity is obtained when $\mathbf{A}(t)$ is not invertible, implying that $\det(\mathbf{A}(t))=0$. We proceed by investigating this case in more detail below.

From Eq. (\ref{ElementsofA}) it can be readily shown that the determinant of the matrix $\mathbf{A}$, suppressing the $t$-dependence, can be written as
\begin{equation}
\begin{aligned}
\det(\mathbf{A})  & = (2\mu B)^3 \Big(\big(\langle X^2\rangle+\langle Y^2\rangle\big)\big(\langle Y^2\rangle\langle X^2\rangle-\langle XY\rangle^2\big)\\
&\quad +\big(\langle Y^2\rangle+\langle Z^2\rangle\big)\big(\langle Y^2\rangle\langle Z^2\rangle-\langle YZ\rangle^2\big)\\
&\quad+\big(\langle Z^2\rangle+\langle X^2\rangle\big)\big(\langle X^2\rangle\langle Z^2\rangle-\langle XZ\rangle^2\big)\\
&\quad +2\big(\langle X^2\rangle\langle Y^2\rangle\langle Z^2\rangle-\langle XY\rangle\langle YZ\rangle  \langle XZ\rangle\big)\Big)\geq 0
\label{DetSymmetricTop}
\end{aligned}
\end{equation}

The Cauchy-Schwarz inequalities between the state vectors $X|\psi(t)\rangle$, $Y|\psi(t)\rangle$, and $Z|\psi(t)\rangle$, which can be expressed in general as
\begin{equation}
    \langle \varphi_1|\varphi_1\rangle\langle\varphi_2|\varphi_2\rangle\geq |\langle\varphi_1|\varphi_2\rangle|^2
    \label{Eq:CS}
\end{equation}
for any two state vectors $|\varphi_1\rangle$ and $|\varphi_2\rangle$, implies that Eq. (\ref{DetSymmetricTop}) is positive semidefinite, as indicated. To see that this holds for the final line in Eq. (\ref{DetSymmetricTop}), we begin with the following relations from Cauchy-Schwarz,
\begin{equation}
    \begin{aligned}
    \langle X^2\rangle\langle Y^2\rangle &\geq \langle XY\rangle^2\\
    \langle Y^2\rangle\langle Z^2\rangle &\geq \langle YZ\rangle^2\\
    \langle Z^2\rangle\langle X^2\rangle &\geq \langle ZX\rangle^2
    \end{aligned}
    \label{Eq:XYZCauchy}
\end{equation}
which may be rearranged by taking products as,
\begin{equation}
\begin{aligned}
     \langle X^2\rangle^2\langle Y^2\rangle^2 \langle Z^2\rangle^2   &\geq  \langle XY\rangle^2\langle YZ\rangle^2\langle ZX\rangle^2 .
\end{aligned}
\end{equation}
Taking the square root of both sides then yields the desired result that 
\begin{equation}
\begin{aligned}
     \langle X^2\rangle\langle Y^2\rangle \langle Z^2\rangle  &\geq  \langle XY\rangle\langle YZ\rangle\langle ZX\rangle.
\end{aligned}
\end{equation}

The equality sign (i.e., a singularity) in Eq. (\ref{DetSymmetricTop}) can arise if and only if $X|\psi(t)\rangle$, $Y|\psi(t)\rangle$, and $Z|\psi(t)\rangle$ are all linearly dependent. The QTC singularity issue will be addressed in Sec. \ref{seciv} below where we describe our
computational methods for solving Eq. (\ref{QTCEquations}).

\section{Computational methods} \label{seciv}

The numerical computation of the QTC fields according to Eq. (\ref{QTCEquations}) requires evaluations of the expectation values for the associated operators. Here, we study QTC of symmetric top molecules in the $|JKM\rangle$ eigenbasis of the drift Hamiltonian, which is given in Eq. (\ref{eq:symtoph0}) and can be rearranged as
\begin{equation}
\begin{aligned}
    H_0 &= B\mathbf{J}^2+(C-B)J_z^2
    \end{aligned}
    \label{H0rearranged}
\end{equation}
leading to the eigenvalue equation
\begin{equation}
\begin{aligned}
H_0\, |JKM\rangle&=\big(BJ(J+1)+(C-B)K^2\big)|JKM\rangle
\end{aligned}
\label{eigenvalueequation}
\end{equation}
where $J=0,1,2,\cdots$ is the total rotational angular momentum quantum number, $K=0,\pm 1,\pm 2,\cdots,\pm J$ is the projection of the angular momentum onto the molecule-fixed $z$-axis, and $M=0,\pm 1,\pm 2,\cdots,\pm J$ is the projection of the angular momentum onto the laboratory frame $Z$-axis. Eq. (\ref{eigenvalueequation}) can be obtained in a straightforward manner from Eq. (\ref{H0rearranged}) using the standard angular momentum matrix element relations $J_z|JKM\rangle = K|JKM\rangle$ and $\mathbf{J}^2|JKM\rangle  = J(J+1)|JKM\rangle$. In this section, we obtain matrix element relations in this basis in order to carry out the two computational steps outlined in Sec. \ref{sec3} that must be taken at each forward time step, i.e., \emph{(i)} solving the time-dependent Schr\"odinger equation, Eq. (\ref{eq:3d1}) and \emph{(ii)} solving the QTC equations, Eq. (\ref{QTCEquations}).

\emph{(i) Solving Eq. (\ref{eq:3d1}):}

We begin by expanding the state of a symmetric top as
\begin{equation}
    |\psi(t)\rangle = \sum_{JKM}\langle JKM|\psi(t)\rangle |JKM\rangle,
\end{equation}
The expansion coefficients are governed by the equation
\begin{equation}
\begin{aligned}
    i&\frac{d}{dt} \langle JKM|\psi(t)\rangle \\
    &= \sum_{J'K'M'}\langle JKM|H_0|J'K'M'\rangle\langle J'K'M'|\psi(t)\rangle \\
    &- \sum_{J'K'M'}\mu\langle JKM|X|J'K'M'\rangle\langle J'K'M'|\psi(t)\rangle \varepsilon_X(t)\\
    &-\sum_{J'K'M'}\mu\langle JKM|Y|J'K'M'\rangle\langle J'K'M'|\psi(t)\rangle \varepsilon_Y(t)\\
    &-\sum_{J'K'M'}\mu\langle JKM|Z|J'K'M'\rangle\langle J'K'M'|\psi(t)\rangle \varepsilon_Z(t),
    \end{aligned}
\end{equation}
where
\begin{equation}
    \langle JKM|H_0|J'K'M'\rangle= BJ(J+1)+(C-B)K^2
\end{equation}
for $J'=J$, $K'=K$, and $M'=M$, and

\begin{widetext}
\begin{equation}
\begin{aligned}
    \langle JKM| X|J'K'M'\rangle 
    &=-  \frac{\mathcal{N}\sqrt{2}(-1)^{2+2J'+M'-K'+2M}}{2}   \sum_{m=-1,1}m \begin{pmatrix}
J &  1 & J'  \\
 M & m & -M'  
 \end{pmatrix} \begin{pmatrix}
  J & 1 & J'  \\
 K & 0 & -K' 
 \end{pmatrix}\,,\\
     \langle JKM| Y|J'K'M'\rangle 
    &=  \frac{\mathcal{N}\sqrt{2}(-1)^{2+2J'+M'-K'+2M}}{2i} \sum_{m=-1,1} \begin{pmatrix}
J &  1 & J'  \\
 M & m & -M'  
 \end{pmatrix} \begin{pmatrix}
  J & 1 & J'  \\
 K & 0 & -K' 
 \end{pmatrix}\,,
    \end{aligned}
        \label{mu_elems}
\end{equation}
with $J'=J\pm 1$, $K'=K$, and $M'=M\pm 1$, and 
\begin{equation}
\begin{aligned}
    \langle JKM| Z|J'K'M'\rangle 
    &= \mathcal{N} (-1)^{2+2J'+M'-K'+2M}   \begin{pmatrix}
J &  1 & J'  \\
 M & 0 & -M'  
 \end{pmatrix} \begin{pmatrix}
  J & 1 & J'  \\
 K & 0 & -K' 
 \end{pmatrix}\,,
    \end{aligned}
        \label{mu_elemsZ}
\end{equation}

\end{widetext}
with $J'=J\pm 1$, $K'=K$, and $M'=M$, in terms of $3j$ symbols, where $\mathcal{N}= \sqrt{(2J+1)(2J'+1)}$ \cite{doi:10.1063/1.1723935,doi:10.1063/1.1727531,Kroto}. The selection rules associated with Eqs. (\ref{mu_elems}) and (\ref{mu_elemsZ}) can be used to accelerate the computation of the associated matrix elements. The selection rules also imply that fields coupling to the system via $X,Y,Z$ can only be used to drive transitions in the quantum numbers $J,M$, while $K$ is conserved.

\emph{(ii) Solving Eq. (\ref{QTCEquations}):}

Eqs. (\ref{mu_elems}) and (\ref{mu_elemsZ}) provide the matrix element relations needed for obtaining the elements of $\mathbf{A}(t)$ in the QTC Eq. (\ref{QTCEquations}) (i.e., see Eq. (\ref{ElementsofA})) in the $|JKM\rangle$ eigenbasis. The computation of $\mathbf{b}(t)$ requires matrix element relations for the triple commutators of the form $[H_0,[H_0,\mathbf{R}]]$, i.e.,
\begin{equation}
\begin{aligned}
   & \langle JKM|[H_0,[H_0,\mathbf{R}]]|J'K'M'\rangle \\
   &= \Big(B\big(J(J+1)-J'(J'+1)\big)\Big)^2\langle JKM|\mathbf{R}|J'K'M'\rangle.
    \end{aligned}
    \label{triplexy}
\end{equation}

The issue of $\det(\mathbf{A}(t))=0$ in Eq. (\ref{DetSymmetricTop}) can be clarified as follows. We will show that the state vectors $X|\psi(t)\rangle$, $Y|\psi(t)\rangle$, and $Z|\psi(t)\rangle$ are linearly independent of each other. Specifically, $X|\psi(t)\rangle$, $Y|\psi(t)\rangle$, and $Z|\psi(t)\rangle$ can be, respectively, further written in terms of the basis $|JKM\rangle$, as
\begin{widetext}
\begin{equation}
\begin{aligned}
&X|\psi(t)\rangle=\sum_{JKM}\Big[\sum_{J'K'M'}\langle JKM|X|J'K'M'\rangle\langle J'K'M'|\psi(t)\rangle \Big] |JKM\rangle, \\
&Y|\psi(t)\rangle=\sum_{JKM}\Big[\sum_{J'K'M'}\langle JKM|Y|J'K'M'\rangle\langle J'K'M'|\psi(t)\rangle \Big]|JKM\rangle, 
\end{aligned}
\end{equation}
and
\begin{equation}
\begin{aligned}
&Z|\psi(t)\rangle=\sum_{JKM}\Big[\sum_{J'K'M'}\langle JKM|Z|J'K'M'\rangle\langle J'K'M'|\psi(t)\rangle \Big]|JKM\rangle, 
\end{aligned}
\end{equation}
\end{widetext}
which, from Eqs. (\ref{mu_elems}) and (\ref{mu_elemsZ}), can be seen to be linearly independent, since the expansion coefficients for $X|\psi(t)\rangle$, $Y|\psi(t)\rangle$, and $Z|\psi(t)\rangle$ in the $|JKM\rangle$ basis are all distinct for bases truncated at some finite, albeit sufficiently large, value $J_{max}$ (which is set to 30 in all of our calculations in Sec. \ref{Sec:Numerics}). As a result, we conclude that $\det(\mathbf{A}(t))>0$ and that singularities will not appear when solving the QTC Eqs. (\ref{QTCEquations}).

\section{Reduction to the case of linear molecules}\label{Sec:LinRotCase}

\begin{figure}[H] 
\centering
\includegraphics[width=0.3\columnwidth]{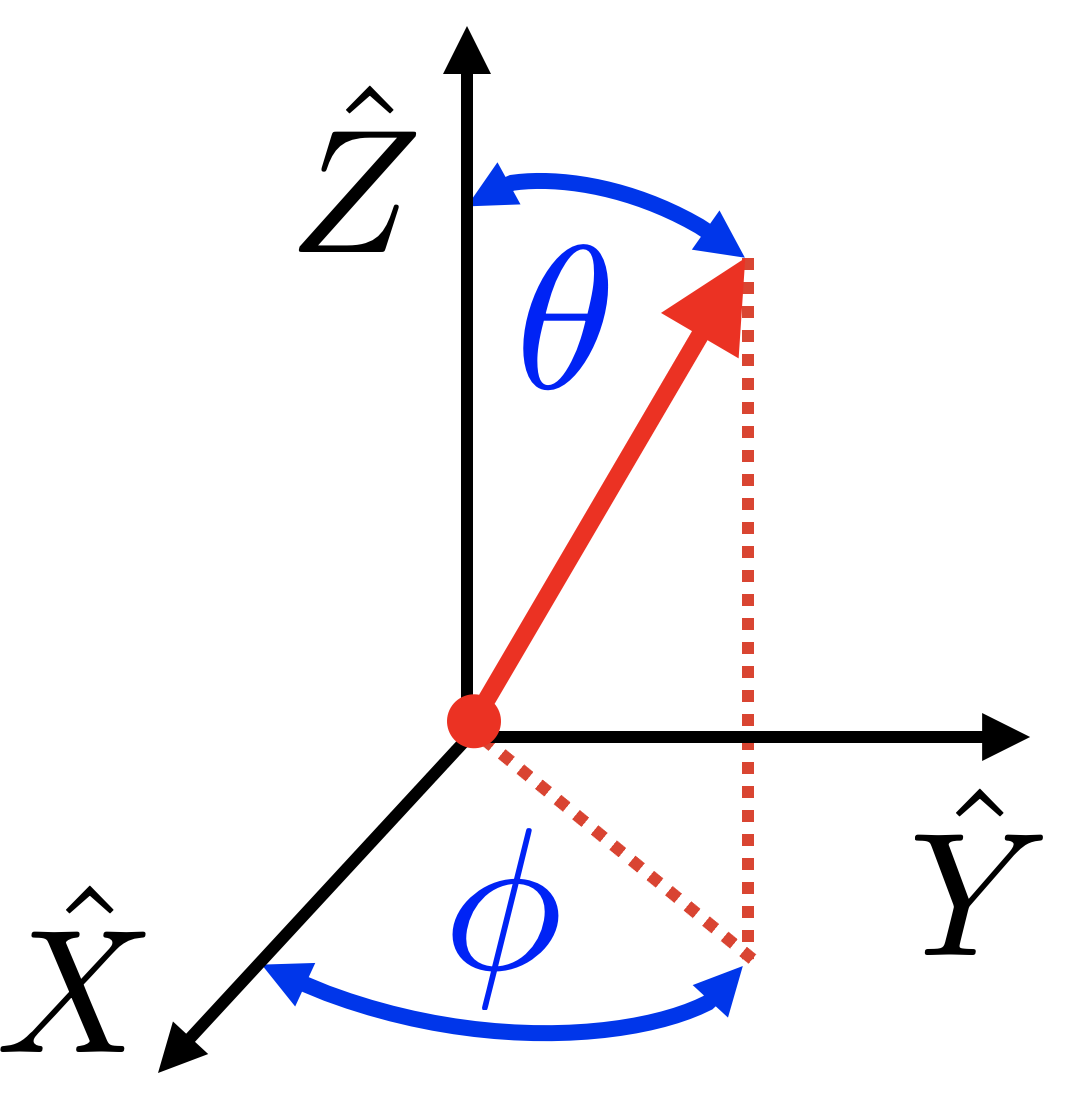}
\caption{Diagram showing $(\theta,\phi)$ relation between laboratory frame fixed $({\hat{X}},{\hat{Y}},{\hat{Z}})$ coordinates (black) and molecular orientation vector (red)}
\label{SphericalPolar}
\end{figure}

Linear molecules possess only one axis of rotation and their Hamiltonian is given by,
\begin{equation}
    H_0 = B\mathbf{L}^2
\end{equation}
where 
\begin{equation}
    \mathbf{L}^2= - \bigg( \frac{1}{\sin\theta}\frac{\partial}{\partial\theta}\Big(\sin\theta\frac{\partial}{\partial\theta}\Big)+\frac{1}{\sin^2\theta}\frac{\partial^2}{\partial\phi^2}\bigg)
    \label{DriftLinear}
\end{equation}
and has no explicit $\chi$-dependence, as depicted in Fig. \ref{SphericalPolar}. This yields an expression for $\det(\mathbf{A})$ that is equal to Eq. (\ref{DetSymmetricTop}). The matrix elements required to study QTC of linear molecules in their eigenbasis can be found using the matrix element relations obtained for symmetric tops and setting $K=0$.

\section{Numerical illustrations}\label{Sec:Numerics}

\begin{figure}[t] 
\centering
\includegraphics[width=1.0\columnwidth]{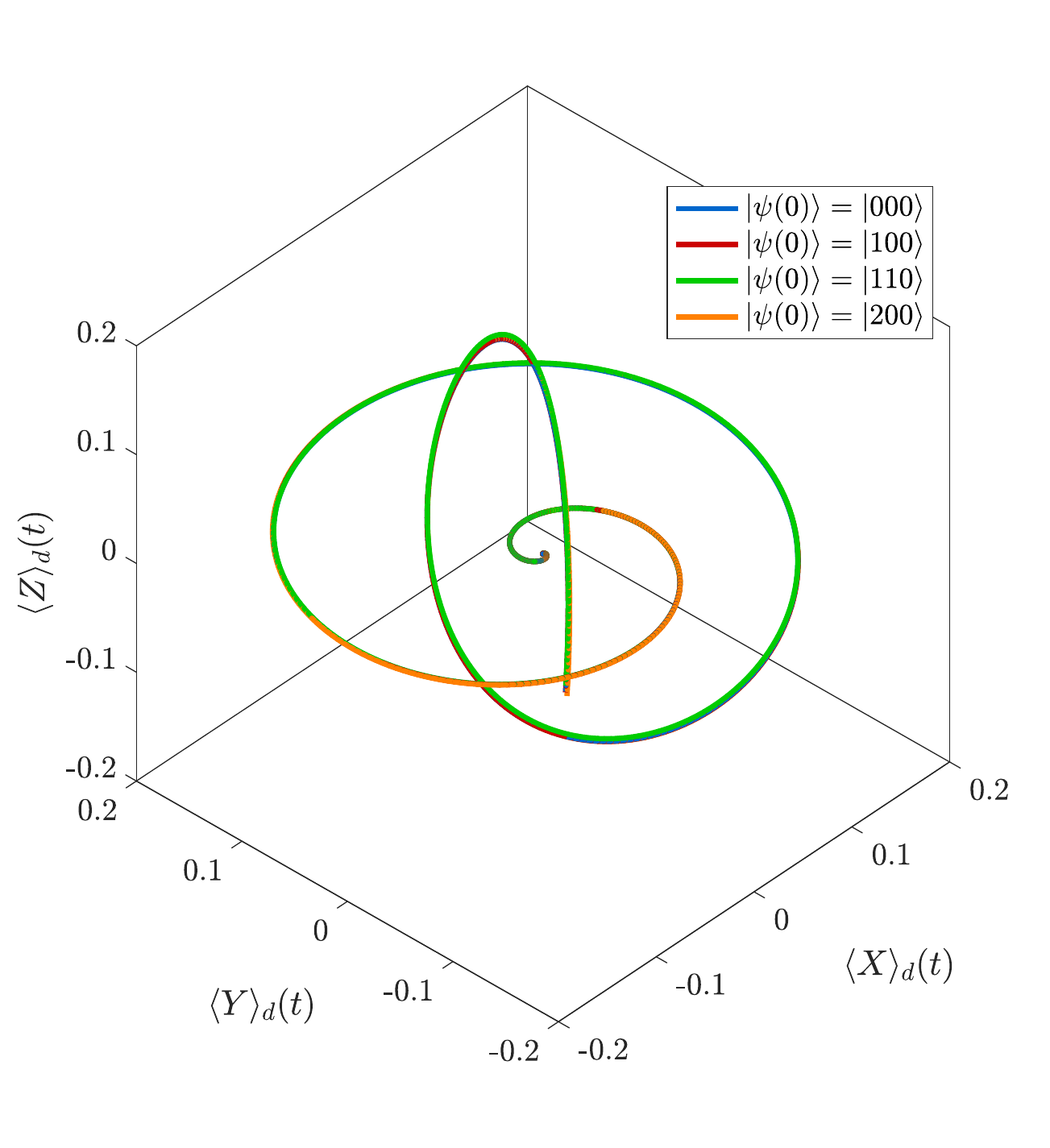}
\caption{The designated tracks $\langle X \rangle_d(t)$, $\langle Y \rangle_d(t)$, and $\langle Z \rangle_d(t)$ given in Eq. (\ref{eq:numericaltracks}) are plotted as a black curve inside of the $\langle X\rangle^2(t) +\langle Y\rangle^2(t)+\langle Z\rangle^2(t)=1$ unit sphere. Then, the QTC tracks $\langle X\rangle(t)$, $\langle Y\rangle(t)$, and $\langle Z\rangle(t)$ followed by the system are plotted in color. The different colors correspond to different initial conditions $|\psi(0)\rangle = |000\rangle,|100\rangle,|110\rangle,|200\rangle$. \label{fig:tracks}}
\end{figure}

\begin{figure}[t] 
\centering
\includegraphics[width=1.0\columnwidth]{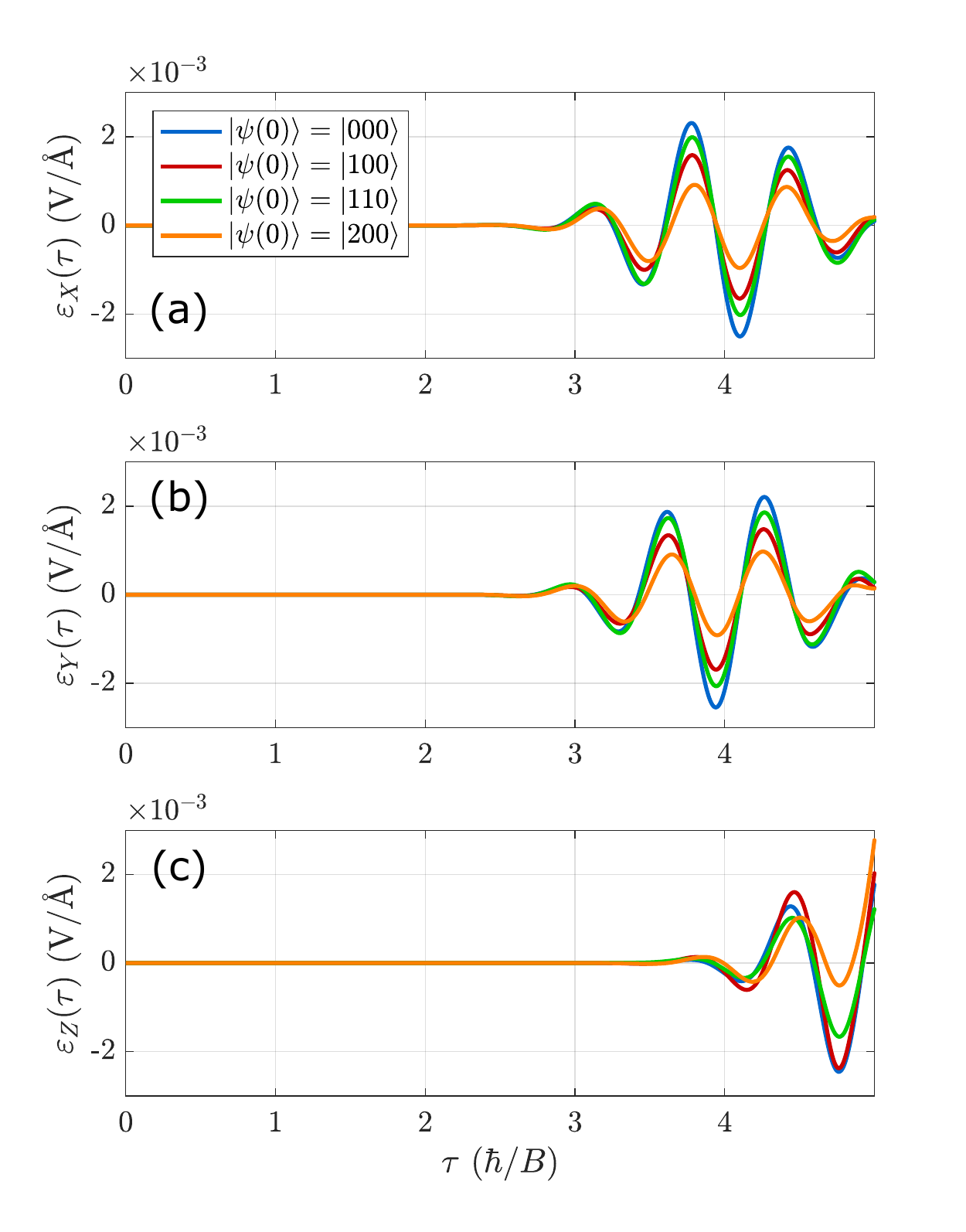} 
\caption{The QTC fields $\varepsilon_X(\tau)$, $\varepsilon_Y(\tau)$, and $\varepsilon_Z(\tau)$ are plotted as a function of the nondimensionalized time $\tau \equiv B t$ in panels (a), (b), and (c), respectively. The different colors correspond to different initial conditions $|\psi(0)\rangle = |000\rangle,|100\rangle,|110\rangle,|200\rangle$.\label{fig:fields} }
\end{figure}

We have derived the QTC equations, Eq. (\ref{QTCEquations}), for controlling symmetric top orientation, and we now present numerical illustrations of this approach. For our illustrations, we consider the symmetric top molecule fluoromethane, with principal rotational constant $B=5.182 \text{ cm}^{-1}$ and second rotational constant $C = 0.852 \text{ cm}^{-1}$ \cite{PAPOUSEK199333}. The magnitude of the dipole moment is given by $\mu=1.847$ Debye \cite{LandoltBornstein1974:sm_lbs_978-3-540-37967-6_1}. The system is represented in the $|JKM\rangle$ basis, with basis elements $|000\rangle$, $\cdots$, $|30,\pm 30,\pm 30\rangle$. We consider designated tracks $\langle X\rangle_d(t)$, $\langle Y\rangle_d(t)$, and $\langle Z\rangle_d(t)$ given by

\begin{equation}
    \begin{aligned}
    \langle X\rangle_d(t)&\equiv 0.2 e^{-\big(\frac{t-0.8T}{T/8}\big)^2}\sin( 8Bt)\\
        \langle Y\rangle_d(t)&\equiv 0.2 e^{-\big(\frac{t-0.8T}{T/8}\big)^2}\cos( 8Bt)\\
            \langle Z\rangle_d(t)&\equiv 0.2 e^{-\big(\frac{t-T}{T/8}\big)^2}\cos( 8Bt)
            \label{eq:numericaltracks}
    \end{aligned}
\end{equation}

where $T=5/B$ is the terminal time and 30,000 time points are used for the calculations. Fig. \ref{fig:tracks} shows a 3D plot comparing these designated $\langle X\rangle_d(t)$, $\langle Y\rangle_d(t)$, and $\langle Z\rangle_d(t)$ trajectories with the actual tracks $\langle X\rangle(t)$, $\langle Y\rangle(t)$, and $\langle Z\rangle(t)$ that are followed when the molecule is initialized in $|\psi(0)\rangle = |000\rangle, |100\rangle,|110\rangle,|200\rangle$. We see that the curves in Fig. \ref{fig:tracks} are all superimposed, indicating that QTC is successful. Meanwhile, Fig. \ref{fig:fields} shows the QTC fields determined via Eq. (\ref{QTCEquations}) that are found to drive $\langle X\rangle(t)$, $\langle Y\rangle(t)$, and $\langle Z\rangle(t)$ along these designated trajectories for the four initial conditions we consider. We note that as per Sec. (\ref{Sec:LinRotCase}), the fields $\varepsilon_X(t)$, $\varepsilon_Y(t)$, and $\varepsilon_Z(t)$ and the tracks associated with $|\psi(0)\rangle = |000\rangle, |100\rangle,|200\rangle$ are the same fields and tracks for a 3D linear rotor with rotational constant $B$, initialized as $|\psi(0)\rangle = |00\rangle, |10\rangle,|20\rangle$.

\section{Conclusions}

In this article, we have explored how QTC can be applied to design fields to orient symmetric top molecules, and have derived expressions for the QTC fields for driving the molecular orientation along time-dependent tracks. We also obtained matrix element relations to facilitate studying QTC of symmetric tops in the $|JKM\rangle$ symmetric top eigenbasis, and presented numerical illustrations of the QTC procedure for driving orientation dynamics in these systems. In order to realize associated experimental demonstrations, molecular rotors could be investigated using, e.g., laser and evaporative cooling methods to create ultracold molecules, and then trapping them in an optical lattice \cite{Baranov2012}. Then, the creation of shaped microwave fields needed for QTC could be explored using arbitrary waveform generators \cite{Yao2011,Lin2005}.

Looking ahead, this QTC formulation could be extended towards studying the control of so-called molecular superrotors \cite{PhysRevLett.112.113004}, e.g. by selecting tracks to create very rapid rotational dynamics. Furthermore, the prospects of applying QTC towards the control of arrays of coupled molecular rotors, e.g. for applications in quantum information science \cite{PhysRevLett.88.067901,PhysRevA.82.062323,doi:10.1002/cphc.201600781}, could be studied as well. For the latter, the study of coupled molecules will likely require high-dimensional modeling to represent the system dynamics, given that the model dimension scales exponentially in the number of degrees of freedom. As such, numerically exact simulations of coupled molecular rotors may not be computationally feasible. However, such challenges may be addressable through the use of suitable approximation frameworks for the quantum dynamics, e.g. \cite{Messina1996,Schroder2008,magann2019quantum,PhysRevLett.106.190501}.

\acknowledgments

A.B.M. acknowledges support from the U.S. Department of Energy, Office of Science, Office of Advanced Scientific Computing Research, Department of Energy Computational Science Graduate Fellowship under Award No. DE-FG02-97ER25308, as well as support from Sandia National Laboratories’ Laboratory Directed Research and Development Program under the Truman Fellowship. H.A.R. acknowledges support from DOE under Grant No.~DE-FG02-02ER15344. T.S.H. acknowledges support from the Army Research Office W911NF-19-1-0382. 

Sandia National Laboratories is a multimission laboratory managed and operated by National Technology \& Engineering Solutions of Sandia, LLC, a wholly owned subsidiary of Honeywell International Inc., for the U.S. Department of Energy's National Nuclear Security Administration under contract DE-NA0003525. This paper describes objective technical results and analysis. Any subjective views or opinions that might be expressed in the paper do not necessarily represent the views of the U.S. Department of Energy or the United States Government. 

This report was prepared as an account of work sponsored by an agency of the United States Government. Neither the United States Government nor any agency thereof, nor any of their employees, makes any warranty, express or implied, or assumes any legal liability or responsibility for the accuracy, completeness, or usefulness of any information, apparatus, product, or process disclosed, or represents that its use would not infringe privately owned rights. Reference herein to any specific commercial product, process, or service by trade name, trademark, manufacturer, or otherwise does not necessarily constitute or imply its endorsement, recommendation, or favoring by the United States Government or any agency thereof. The views and opinions of authors expressed herein do not necessarily state or reflect those of the United States Government or any agency thereof. 

\bibliography{bib}

 \end{document}